\documentclass[floatfix]{revtex4}
\usepackage[dvips]{graphicx}
\usepackage{fancyhdr}
\usepackage{palatino,epsfig}
\begin{document}

\title{LHAPDF: PDF Use from the Tevatron to the LHC}

\author{ D Bourilkov$^\ddagger$, R C Group$^\ddagger$\footnote{This talk was given by R. C. Group in October 2005 at the TeV4LHC workshop, describing ongoing work performed in collaboration with the other authors.}, and M R Whalley$^\dagger$,}
\affiliation{
University of Florida$^\ddagger$\\
Gainesville, FL 32611, USA \\
and\\
University of Durham$^\dagger$\\
Durham, DH1 3LE, UK\\
}


\begin{abstract}
Parton Density Functions (PDFs) and their uncertainties are extremely important topics for both the Tevatron and the LHC.  Experiments at the Tevatron can enhance this knowledge not only by constraining the PDF fits, but also by developing and refining the available PDF tools through feed-back from the experiments that are currently analyzing the highest energy hadron collider data available.  It is important that the community has standardized tools and methods at its disposal.  In this note we summarize briefly the most recent developments of the The \underline{L}es \underline{H}ouches \underline{A}ccord \underline{PDF} (LHAPDF), which is the modern replacement for PDFLIB.  We also outline and compare the methods of quantifying the impact of PDF uncertainties on physical observables.  The PDF weighting method for propagating errors from PDFs to event generator observables is outlined in detail, and example code for using this method with {\tt PYTHIA} is also included.
\end{abstract}

\maketitle

\thispagestyle{fancy}


\section{Introduction}
The experimental errors in current and future hadron colliders are expected to decrease to a level that will challenge the uncertainties in theoretical calculations. One important component in the prediction of uncertainties at hadron colliders comes from the Parton Density Functions (PDFs) of the (anti)proton.  

 The highest energy particle colliders in the world currently, and in the near future, collide hadrons.  To make predictions of hadron collisions, the parton cross sections must be folded with the parton density functions:

\begin{equation}
\rm \frac{d\ \sigma}{d\; variable}[pp\rightarrow X] \sim \sum_{ij} \, \left(f_{i/p}(x_1)  f_{j/p}(x_2) +(i \leftrightarrow j)\right)\,  \hat{\sigma}  
\label{llpdf1}
\end{equation}

$\rm \hat{\sigma}$ - cross section for the partonic subprocess $ij\rightarrow X$\\
$\rm x_1$, $\rm x_2$ - parton momentum fractions,\\
$\rm f_{i/p(\bar{p})}(x_i)$ - probability to find a parton $i$ with momentum
fraction $x_i$ in the (anti)proton.\\

A long standing problem when performing such calculations is to quantify the uncertainty of the results coming from our limited knowledge of the PDFs.  Even if the parton cross section $\rm \hat{\sigma}$ is known very precisely, there may be a sizable error on the hadronic cross section $\sigma$
due to the PDF uncertainty.

The Tevatron can contribute to PDF knowledge in many ways that will benefit the experiments at the LHC.  First, measurements made by the experiments at FNAL will reduce PDF uncertainties by constraining PDF fits.  Perhaps more importantly tools and techniques for propagating PDF uncertainty through to physical observables can be improved and tested at the Tevatron.  

Next-to-leading order (NLO) is the first order at which the the normalization of the hard-scattering cross sections has a reasonable uncertainty.  Therefore, this is the first order at which PDF uncertainties are usually applied.  To date, all PDF uncertainties have been calculated in the context of NLO global analysis.  However, useful information can still be obtained from NLO PDF uncertainties with leading order (LO) calculations and parton shower Monte Carlos~\cite{Joey}.  

Techniques and tools for calculating PDF uncertainty in the context of LO parton shower Monte Carlos will be the primary topic of this document.  Examples are provided employing CTEQ6~\cite{Pumplin:2002vw} error sets from LHAPDF and the parton shower Monte Carlo program {\tt PYTHIA}~\cite{PYTHIA}.

\section{LHAPDF update}

\subsection{Recent developments}

Historically, the CERN PDFLIB library~\cite{pdflib} has provided a widely used standard FORTRAN interface to PDFs with interpolation grids built into the PDFLIB code itself. However, it was realized that PDFLIB would be increasingly unable to meet the needs of the new generation of PDFs which often involve large numbers of sets ($\approx$20--40) describing the uncertainties on the individual partons from  variations in the fitted parameters.  As a consequence of this, at the Les Houches meeting in 2001~\cite{giele:2002hx}, the beginnings of a new interface were conceived --- the so-called \underline{L}es \underline{H}ouches \underline{A}ccord \underline{PDF} (LHAPDF).  The LHAGLUE package~\cite{Bourilkov:2003kk} plus a unique PDF numbering scheme enables LHAPDF to be used in the same way as PDFLIB, without requiring {\em any} changes in the {\tt PYTHIA} or {\tt HERWIG} codes.  The evolution of LHAPDF (and LHAGLUE) up to summer 2005 is well documented~\cite{Whalley:2005nh}.  
  
Recently, LHAPDF has been further improved.  With the release of v4.1 in August of 2005 the installation method has been upgraded to the more conventional {\tt configure; make; make install}.  Version 4.2, released in November of 2005, includes the  new cteq6AB (variable $\alpha(M_Z)$) PDF sets.  It also includes new modifications by the CTEQ group to other cteq code to improve speed.   Some minor bugs were also fixed in this version that affected the a02m$\_$nnlo.LHgrid file (previous one was erroneously the same as LO) and SMRSPI code which was wrongly setting {\tt usea} to zero.    

A v5 version, with the addition of the option to store PDFs from multiple sets in memory, has been released.  This new functionality speeds up the code by making it possible to store PDF results from many sets while only generating a MC sample once without significant loss of speed.

\subsection{Consistency checks}
As a technical check, cross sections have been computed, as well as errors where appropriate, for all PDF sets included in LHAPDF. 10,000 events are generated
for each member of a PDF set for both {\tt HERWIG}~\cite{HERWIG} and {\tt PYTHIA}~\cite{PYTHIA},
and at both Tevatron and LHC energies.
As this study serves simply as a technical check of the
interface, no attempt was made to unfold the true PDF error.
The maximum Monte Carlo variance (integration error) in our checks is less than 1 percent.
This has not been subtracted and will result in an overestimate of the true
PDF uncertainty by a factor $<\sim 1.05$ in our analysis.  The results in general show good agreement for most PDFs included in the checks. Overall the consistency is better for Tevatron energies, where we do not have to make large extrapolations to the new energy domain and much broader phase space covered by the LHC.

Two complementary processes are used:
\begin{itemize}
\item Drell--Yan Pairs ($\mu^+\mu^-$):
the Drell--Yan process is chosen here to probe the functionality of the
quark PDFs included in the LHAPDF package.  
\item Higgs Production:
the cross section for $gg \rightarrow H$ probes the gluon PDFs, so this
channel is complementary to the case considered above.
\end{itemize}

\section{PDF uncertainties}
As stated above, the need to understand and reduce PDF uncertainties in theoretical predictions for collider physics is of paramount importance.  One of the first signs of this necessity was the apparent surplus of high $P_T$ events observed in the inclusive jet cross section in the CDF experiment at FNAL in run I.  Subsequent analysis of the PDF uncertainty in this kinematic region indicated that the deviation was within the range of the PDF dominated theoretical uncertainty on the cross section.  Indeed, when the full jet data from the Tevatron (including the D0 measurement over the full rapidity range) was included in the global PDF analysis, the enhanced high x gluon preferred by CDF jet data from Run I became the central solution.  This was an overwhelming sign that PDF uncertainty needed to be quantified~\cite{Giele:2001mr}. Below, a short review of one approach to quantify these uncertainties called the Hessian matrix method is given, followed by outlines of two methods used to calculate the PDF uncertainty on physical observables. 

\subsection{Review of the Hessian Method}
 Experimental constraints must be incorporated into the uncertainties of parton distribution functions before these uncertainties can be propagated through to predictions of observables.  The Hessian Method~\cite{Pumplin:2001} both constructs a N Eigenvector Basis of PDFs and provides a method from which uncertainties on observables can be calculated.  
 The first step of the Hessian method is to make a fit to data using N free parameters.  The global $\chi ^2$ of this fit is minimized yielding a central or best fit parameter set $S_0$.  Next the global  $\chi ^2$ is increased to form the Hessian error matrix:
\begin{equation}
\Delta \chi ^{2}=\sum_{i=1}^N \sum_{j=1}^N H_{ij}(a_{i}-a_{i}^{0})(a_{j}-a_{j}^{0})
\end{equation}

This matrix can then be diagonalized yielding N (20 for CTEQ6) eigenvectors.  Each eigenvector probes a direction in PDF parameter space that is a combination of the 20 free parameters used in the global fit.  The largest eigenvalues correspond to the best determined directions and the smallest eigenvalues to the worst determined directions in PDF parameter space.  For the CTEQ6 error PDF set, there is a factor of roughly one million between the largest and smallest eigenvectors.  The eigenvectors are numbered from highest eigenvalue to lowest eigenvalue.  Each N eigenvector direction is then varied up and down within tolerance to obtain 2N new parameter sets, $S^{\pm}_i(i=1,..,N)$.  These parameter sets each correspond to a member of the PDF set, $F_{i}^{\pm}=F(x,Q;S_{i}^{\pm})$. The PDF library described above, LHAPDF, provides standard access to these PDF sets.

\subsection{'Master' Equations}
  Although the variations applied in the eigenvector directions are symmetric by construction, this is not always the case for the result of these variations when propagated through to an observable.  In general the well constrained directions (low eigenvector numbers) tend to have symmetric positive and negative deviations on either side of the central value of the observable (X$_0$).  This can not be counted on in the case of the smaller eigenvalues (larger eigenvector numbers).  The 2N+1 members of the PDF set provide 2N+1 results for any observable of interest.  Two methods for obtaining a set of results are described in detail below.  Once results are obtained they can be used to approximate PDF uncertainty through the use of a 'Master Equation'.  Although many versions of these equations can be found in the literature, the type which considers maximal positive and negative variations of the physical observable separately, known as ``modified tolerance method'', is preferred~\cite{Sullivan:2002jt,Sullivan}:

\begin{equation}  
\Delta X^{+}_{max}=\sqrt{\sum_{i=1}^N[max(X^{+}_{i}-X_{0},X^{-}_{i}-X_{0},0)]^{2}}
\end{equation}

\begin{equation}  
\Delta X^{-}_{max}=\sqrt{\sum_{i=1}^N[max(X_{0}-X^{+}_{i},X_{0}-X^{-}_{i},0)]^{2}}
\end{equation}  

Other forms of 'Master' equations with their flaws are summarized:
\begin{itemize}

\item 
$\Delta X_{1}=\frac{1}{2}\sqrt{\sum_{i=1}^N  (X^{+}_{i}-X^{-}_{i})^{2}}$ \\
This is the original CTEQ 'Master Formula'.  It correctly predicts uncertainty on the PDF values since in the PDF basis $X^{+}_{i}$ and  $X^{-}_{i}$ are symmetric by construction.  However, for physical observables this equation will underestimate the uncertainty if $X^{+}_{i}$ and  $X^{-}_{i}$ lie on the same side of $X_{0}$.\\

\item
 $\Delta X_{2}=\frac{1}{2}\sqrt{\sum_{i=1}^{2N}  R_i^{2}}$ ($R_{1}= X^{+}_{1}-X_{0}$,$R_{2}= X^{-}_{1}-X_{0}$, $R_{3}= X^{+}_{2}-X_{0}$ ...)\\
If $X^{+}_{i}$ and  $X^{-}_{i}$ lie on the same side of $X_{0}$ this equation adds contributions from both in quadrature.  NOTE: For symmetric and asymmetric deviations, $\Delta X_{1}$ varies from $0\rightarrow \frac{2}{\sqrt{2}} \Delta X_{2}$\\

\item  positive and negative variations based on eigenvector directions\\
 $\Delta X^{+}=\sqrt{\sum_{i=1}^N(X^{+}_{i}-X_{0})^{2}}$,\ \ $\Delta X^{-}=\sqrt{\sum_{i=1}^N(X^{-}_{i}-X_{0})^{2}}$\\
Since the positive and negative directions defined in the PDF eigenvector space are not always related to positive and negative variations on an observable these equations can not be interpreted as positive and negative errors in the general case.

\end{itemize}

\subsection{Summary of Techniques}
Two main techniques are currently employed to study the effect of PDF uncertainties of physical observables.  Both techniques work with the PDF sets derived from the Hessian method.   

\subsubsection{'Brute Force'}
The 'brute force' method simply entails running the MC and obtaining the observable of interest for each PDF in the PDF set.  This method is robust, and theoretically correct.  Unfortunately,  it can require very large CPU time since large statistical samples must be generated in order for the PDF uncertainty to be isolated over statistical variations.  This method generally is unrealistic when detector simulation is desired.   

Because the effect on the uncertainty of the PDF set members is added in quadrature, the uncertainty is often dominated by only a few members of the error set.  In this case, a variation of the 'brute force' method can be applied.  Once the eigenvectors that the observable is most sensitive to are determined, MC samples only need to be generated for the members corresponding to the variation of these eigenvalues.  This method will always slightly underestimate the true uncertainty.

\subsubsection{PDF Weights}
As mentioned above, often it is not possible to generate the desired MC sample many times in order to obtain the uncertainty on the observable due to the PDF.  The 'PDF Weights' method solves this problem~\cite{Joey}.  The idea is that the PDF contribution to Equation~\ref{llpdf1} may be factored out.  That is, for each event generated with the central PDF from the set, a PDF weight ($W_{n}^{0}=1,W_{n}^{i}=\frac{f(x_{1},Q;S_{i})f(x_{2},Q;S_{i})}{f(x_{1},Q;S_{0})f(x_{2},Q;S_{0})}$ where $n=1...N_{events},i=1..N_{PDF}$) can be stored for each event.  The PDF weight technique can be summarized as follows... 

\begin{itemize}
\item Only one MC sample is generated but 2N (e.g. 40) PDF weights are obtained for 
\begin{equation} 
W_{n}^{0}=1,W_{n}^{i}=\frac{f(x_{1},Q;S_{i})f(x_{2},Q;S_{i})}{f(x_{1},Q;S_{0})f(x_{2},Q;S_{0})}
\end{equation}

 where $n=1...N_{events},i=1..N_{PDF}$

\item Only one run, so kinematics do not change and there is no residual statistical variation in uncertainty.
\item The observable must be weighted on an event by event basis for each PDF of the set.  One can either store a ntuple of weights to be used 'offline', or fill a set of weighted histograms (one for each PDF in the set).
\end{itemize}

The benefits of the weighting technique are twofold.  First, only one sample of MC must be generated.  Second, since the observable for each PDF member is obtained from the same MC sample there is no residual statistical fluctuation in the estimate of the PDF uncertainty.  One concern involving this method is that re-weighting events does not correctly modify the Sudakov form factors.  However, the difference in this effect due to varying the PDF was shown to be negligible ~\cite{Gieseke}.  That is, the initial state parton shower created with the central PDF (CTEQ6.1) also accurately represents the parton shower that would be produced by any other PDF in the error set.
 
The weighting method is only theoretically correct in the limit that all possible initial states are populated.  For this reason, it is important that reasonable statistical samples are generated when using this technique.  Any analysis which is sensitive to the extreme tails of distributions should use this method with caution. 

There are two options for using the PDF weighting technique.  One can either store 2N (e.g. 40 for CTEQ) weights for each event, or store  $X_{1},X_{2},F_{1},F_{2}$, and $Q^{2}$ and calculate the weights 'offline'.  The momentum of the two incoming partons may be obtained from {\tt PYTHIA} via  PARI(33) and PARI(34).  Flavour types of the 2 initial partons are stored in $F_{1}=MSTI(15)$ and $F_{2}=MSTI(16)$, and the numbering scheme is the same as the one used by LHAPDF, Table~\ref{flavour-scheme}, except that the gluon is labeled '21' rather than '0'.  The $Q^2$ of the interaction is stored in  $Q^{2}=PARI(24)$.  In theory, this information and access to LHAPDF is all that is needed to use the PDF weights method. This approach has the additional benefit of enabling the 'offline reweighting' with new PDF sets, which have not been used, or even existed, during the MC generation.   We plan to include sample code facilitating the use of PDF weights in future releases of LHAPDF.

\begin{table}[ht]
\begin{center}
\caption{The flavour enumeration scheme used for {\sl f(n)} in LHAPDF}
\label{flavour-scheme}
\begin{tabular}{
|p{1.0cm}
*{13}{|p{0.4cm}}
|}
\hline
parton &
$\bar{t}$ &
$\bar{b}$ &
$\bar{c}$ &
$\bar{d}$ &
$\bar{u}$ &
$\bar{d}$ &
g &
d &
u &
s &
c &
b &
t \\
\hline
{\sl n} & -6 & -5 & -4 & -3 & -2 & -1 & 0 & 1 & 2 & 3 & 4 & 5 & 6 \\
\hline
\end{tabular}
\end{center}
\end{table}

\section{Example Studies}

\subsection{Drell--Yan at the LHC}

The Drell--Yan process is chosen as an almost ideal test case involving quark PDFs for the different flavours.

\begin{figure*}[htb]
\includegraphics[width=80mm]{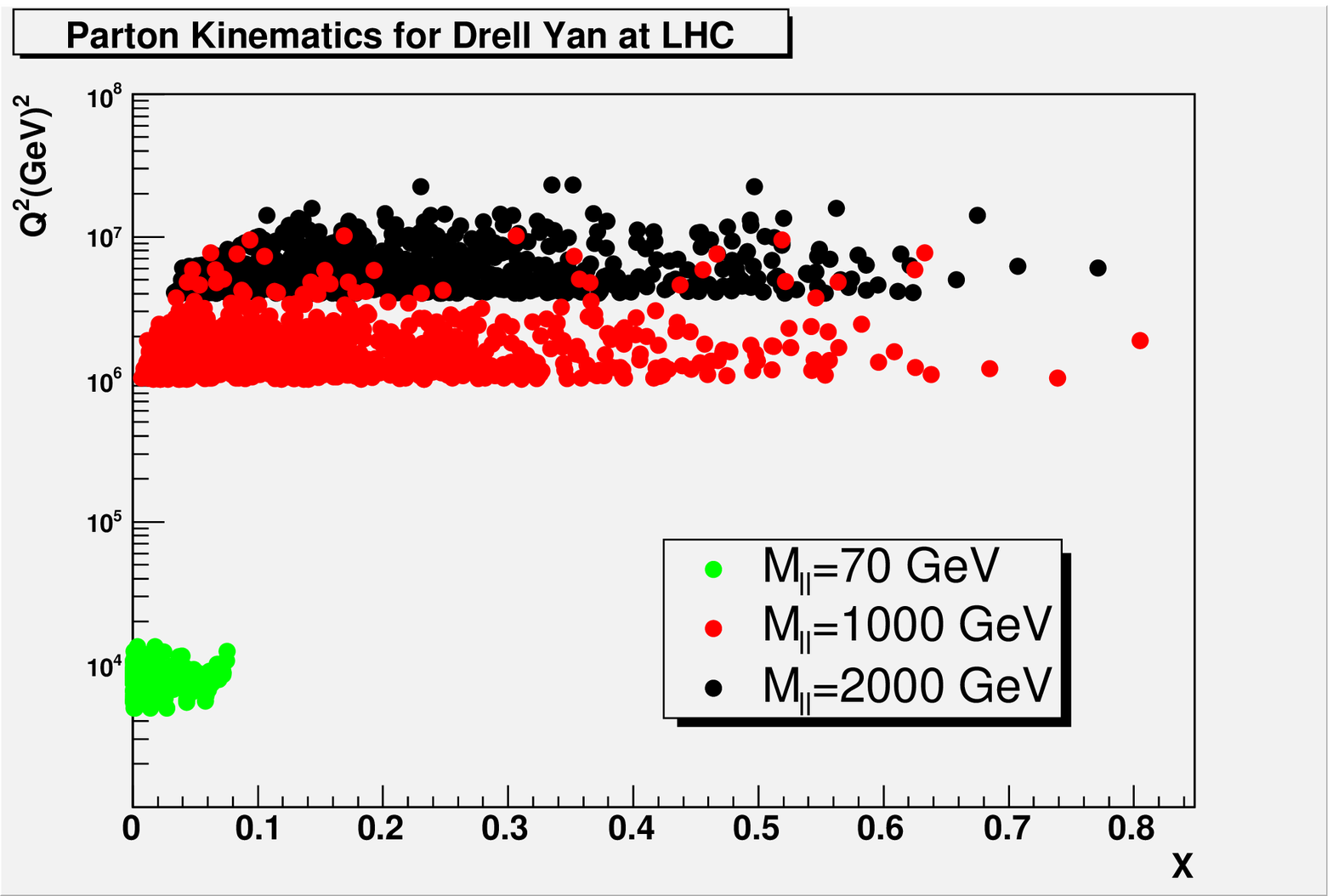} 
\includegraphics[width=80mm]{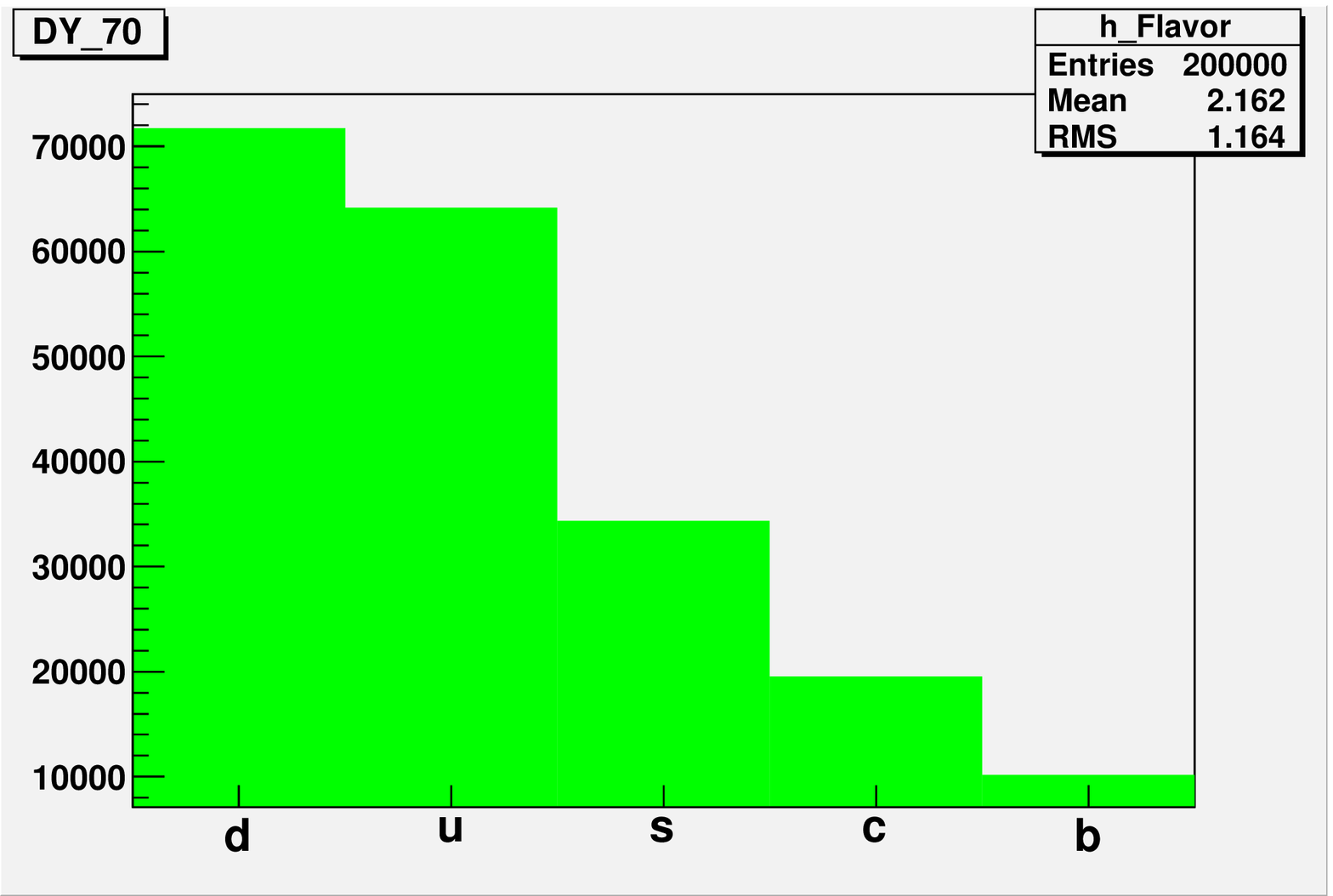} 
\includegraphics[width=80mm]{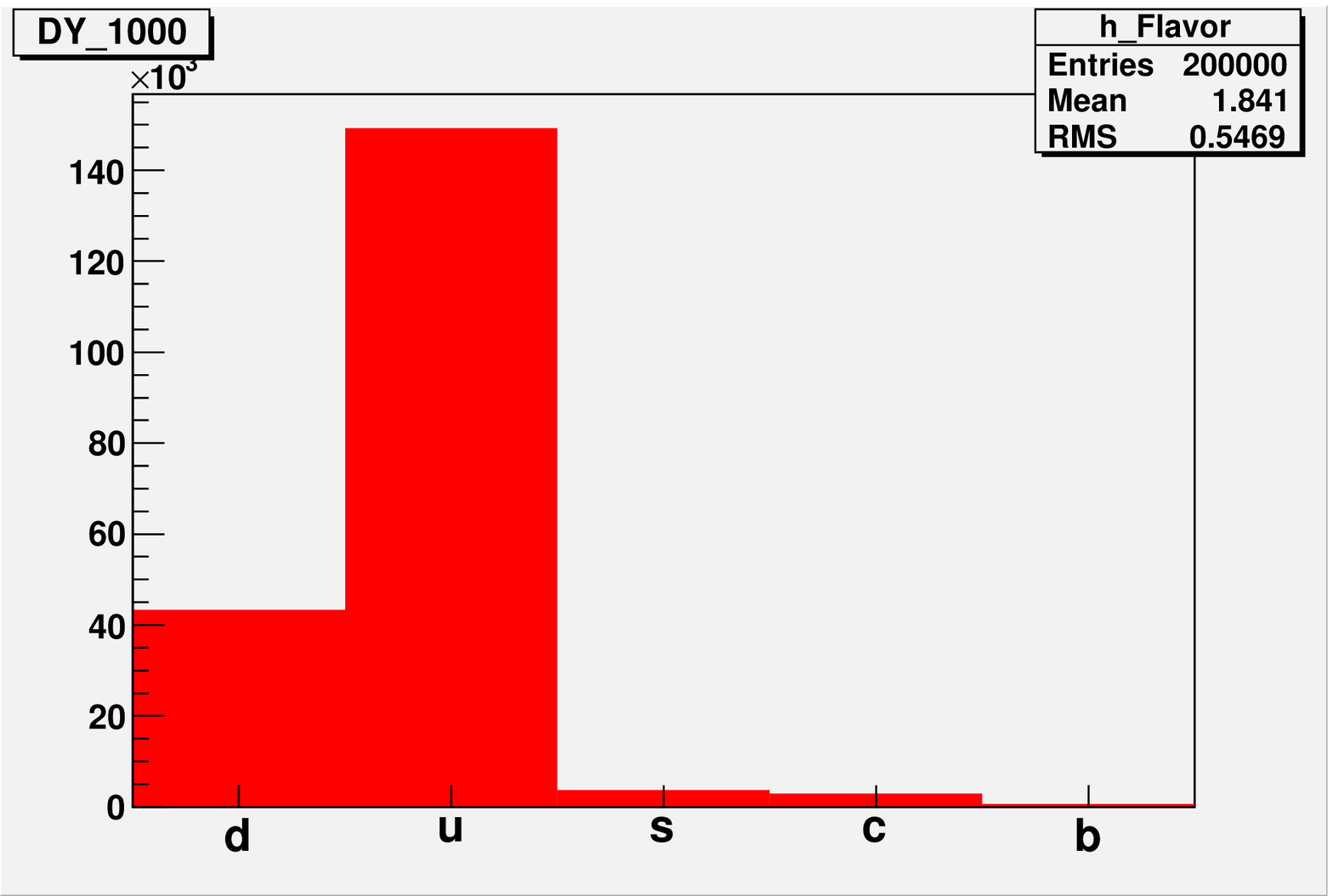} 
\includegraphics[width=80mm]{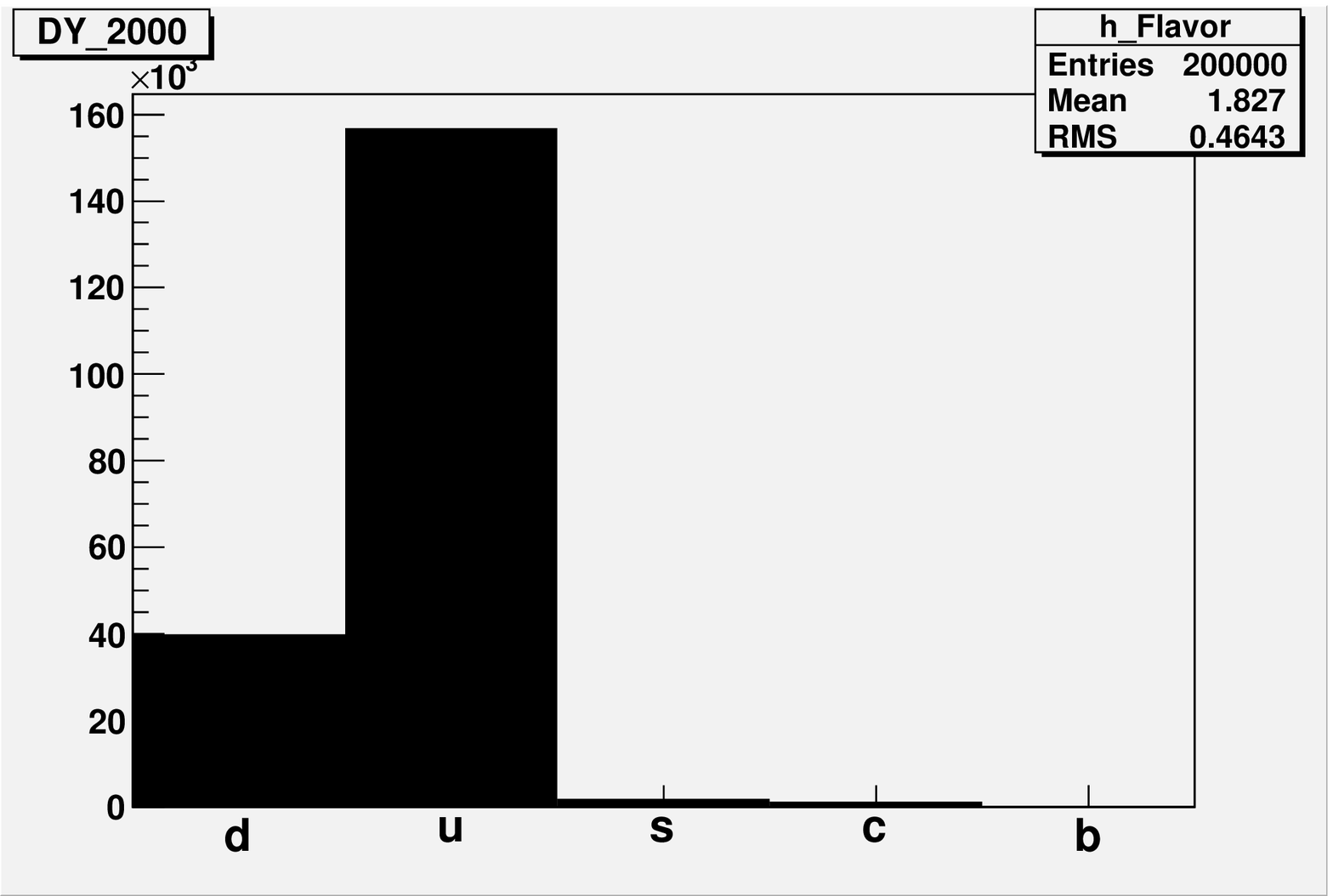} 
\caption{The parton kinematics for Drell--Yan production at the LHC for three Drell--Yan mass choices.  Also the initial parton flavour content for the three cases is shown.}
\label{DY_figs}
\end{figure*}
The initial state parton kinematics and flavour contributions are given in Figure~\ref{DY_figs} for three regions of invariant mass of the final state lepton pair: $70 < M < 120,\ M > 1000,\ M > 2000$~GeV. As we can observe, they cover very wide range in X and Q$^2$. It is interesting to note that the flavour composition around the Z peak contains important contributions from five flavours, while at high mass the u and d quarks (in ratio 4:1) dominate almost completely.

\subsection{Higgs Production in gg$\rightarrow$H at the LHC}
This channel is chosen as complementary to the first one and contains only contributions from the gluon PDF. A light Higgs mass of 120 GeV is selected.

\subsection{Inclusive Jet Cross Section at the Tevatron and the LHC}
\begin{figure*}[htb]
\centering
\includegraphics[width=80mm]{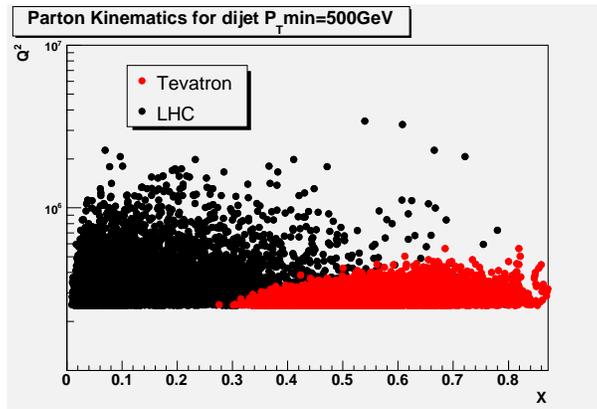} \\
\caption{The partonic jet kinematics for the inclusive jet cross section at the Tevatron and the LHC.}
\label{jet_kinematics}
\end{figure*}

As mentioned above, the inclusive jet cross section was one of the first measurements where the need to quantify PDF uncertainty was evident.  QCD 2-2 processes are studied for $\hat{P_T}\ >\ 500\ GeV$.  The kinematic range probed can be seen in Figure~\ref{jet_kinematics}.

\subsection{Results}
The results for all 3 studies are summarized in Table~\ref{result_table}.
The weighting technique produces the same results as the more elaborate
'brute force' approach for all cases.

\begin{table}[htb]
{
\normalsize
\begin{center}
\caption{Results for the 3 case studies. The central values of the
cross sections in [pb] are shown, followed by the estimates of the
uncertainties for the different master equations and the 'brute force' (B.F.)
and weighting (W) techniques.}
\label{result_table}
\begin{tabular}{|l|c|c|c|cc||cc|}

\hline
Process (method)          & $X_{0}$(best fit)   &  $\Delta X_{1}$ & $\Delta X_{2}$ &$+\Delta X^{+}$& $-\Delta X^{-} $ & $+\Delta X_{max}^{+}$& $-\Delta X_{max}^{+}$ \\ \hline
DY 70$<$M$<$110 (B.F.)   & 1086      &48   & 42  & +55&-63   &+51&-62 \\ \hline
DY 70$<$M$<$110 (W)     & 1086      &48   & 42  & +55&-64    &+51&-63 \\ \hline \hline
DY M$>$1000 (B.F.)  & 6.7e-3      & 3.5e-4   & 2.6e-4  & + 3.5e-4&-3.8e-4   &+3.4e-4&-3.9e-4 \\ \hline
DY M$>$1000 (W)     & 6.7e-3      &3.5e-4.   & 2.6e-4  & +3.5e-4&-3.8e-4    &+3.4e-4&-3.8e-4 \\ \hline \hline
DY M$>$2000 (B.F.)  & 2.2e-4      &1.8e-5   & 1.3e-5  & +1.9e-5& -1.9e-5   &+2.0e-5&-1.7e-5 \\ \hline
DY M$>$2000 (W)     & 2.2e-4      & 1.8e-5   & 1.3e-5  & +1.8e-5&-1.9e-5    &+2.0e-5&-1.7e-5 \\ \hline\hline\hline 
$gg\rightarrow H$ (B.F.)  & 17      &.94   & .68  & +.82&-1.1   &+.8&-1.1 \\ \hline
$gg\rightarrow H$ (W)     & 17      &.94   & .68  & +.82&-1.1    &+.8&-1.1 \\ \hline \hline \hline
DJ500 TeV (B.F.)  & 0.022        & 0.0068   & 0.0057     &0.0048   &0.010    &0.011      &0.0042  \\ \hline
DJ500 TeV (W)     & 0.022        & 0.0068   & 0.0057     &0.0048   &0.010    &0.011      &0.0042  \\ \hline\hline

DJ500 LHC (B.F.)  & 880         &63       & 47      &56      &74      &76        &53    \\ \hline
DJ500 LHC (W)     & 880         &63       & 47      &57      &75      &77        &53    \\ \hline

    \end{tabular}
\end{center}
}
\end{table}

\section{Summary}
In this talk new developments of LHAPDF and consistency checks for all PDF sets are described. The approaches to PDF uncertainty analysis are outlined and the modern method of PDF weighting is described in detail and tested in different channels of current interest.  Drell-Yan, gluon fusion to Higgs, and high $P_T$ jet production are studied at the Tevatron and LHC energy scales.  The methods are in agreement in all cases.  Equations for quantifying PDF uncertainty are discussed and the type which relies on maximal positive and negative variations on the observable is considered superior.

\subsection*{Acknowledgments}

The authors would like to thank Joey Huston for encouraging this study.
DB wishes to thank the United States National Science Foundation
for support from grant NSF 0427110 (UltraLight). CG wishes to thank the US
Department of Energy for support from an Outstanding Junior Investigator award
under grant DE-FG02-97ER41209. MRW wishes to thank the UK PPARC for support from
grant PP/B500590/1.


{\raggedright

\bibliography{TeV4LHC_PDF}
}
\end{document}